\documentclass[sigconf]{acmart}

\setcopyright{none}
\renewcommand\footnotetextcopyrightpermission[1]{}

\usepackage{booktabs}
\usepackage{float}
\usepackage{graphicx}
\usepackage{xcolor}
\usepackage{tikz}
\usetikzlibrary{positioning, arrows.meta, fit, backgrounds}

\begin{document}

\title{PSI: Shared State as the Missing Layer for Coherent AI-Generated Instruments in Personal AI Agents}

  \author{Zhiyuan Wang}
  \affiliation{%
    \institution{University of Virginia}
    \city{Charlottesville}
    \state{Virginia}
    \country{USA}
  }
  \email{vmf9pr@virginia.edu}

  \author{Erzhen Hu}
  \affiliation{%
    \institution{University of Virginia}
    \city{Charlottesville}
    \state{Virginia}
    \country{USA}
  }
  \email{eh2qs@virginia.edu}

  \author{Mark Rucker}
  \affiliation{%
    \institution{University of Virginia}
    \city{Charlottesville}
    \state{Virginia}
    \country{USA}
  }
  \email{mr2an@virginia.edu}

  \author{Laura Barnes}
  \affiliation{%
    \institution{University of Virginia}
    \city{Charlottesville}
    \state{Virginia}
    \country{USA}
  }
  \email{lb3dp@virginia.edu}

% ============================================================
\begin{abstract}
Personal AI tools can now be generated from natural-language requests, but they often remain isolated after creation. We present PSI, a shared-state architecture that turns independently generated modules into coherent instruments: persistent, connected, and chat-complementary artifacts accessible through both GUIs and a generic chat agent. By publishing current state and write-back affordances to a shared personal-context bus, modules enable cross-module reasoning and synchronized actions across interfaces.
We study PSI through a three-week autobiographical deployment in a self-developed personal AI environment and show that later-generated instruments can be integrated automatically through the same contract.
PSI identifies shared state as the missing systems layer that transforms AI-generated personal software from isolated apps into coherent personal computing environments.

\end{abstract}




\begin{teaserfigure}
\centering
\includegraphics[width=0.9\textwidth]{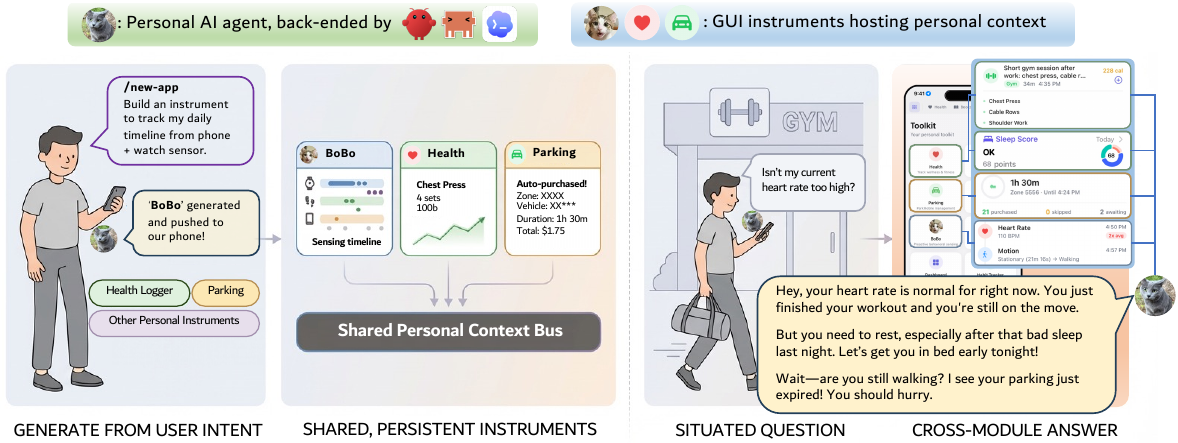}
\caption{\textbf{PSI at a glance.} (1) A user describes a personal need in natural language; an AI generation engine produces a shared-state instrument. (2) Each module publishes structured state to a shared personal-context bus. (3) Both a chat agent and persistent GUIs read from the same bus. (4) This shared state enables grounded, cross-module reasoning and bidirectional actions across the personal computing environment.}
\Description{A wide four-panel horizontal storyboard illustrating the PSI workflow. Panel 1 (Generate from User Intent): A person asks an AI agent to build a 'BoBo' instrument to track their daily timeline using phone and watch sensor data. Panel 2 (Shared, Persistent Instruments): Three persistent GUI cards—BoBo (sensing timeline), Health (workout logs), and Parking (status)—are shown publishing their data into a 'Shared Personal Context Bus'. Panel 3 (Situated Question): The user, walking away from a gym, asks their phone if their current heart rate of 100 is too high. Panel 4 (Cross-Module Answer): The AI provides a grounded response explaining that the heart rate is normal given the recent workout, current walking, and poor sleep, while also issuing a proactive alert that their parking has expired.
}
% \caption{PSI architecture overview. A user describes a personal need in natural language; an AI generation engine produces a module bundle (GUI, service, context provider, action endpoints). Each module publishes structured state to a shared personal-context bus. Both the chat agent and persistent GUIs read the same bus; actions write back through it.}
% \Description{Architecture diagram showing the PSI pipeline from natural-language need through AI generation to module bundles, a shared personal-context bus, and dual-modality consumption by chat and GUI interfaces.}
\label{fig:teaser}
\end{teaserfigure}

\maketitle

% ============================================================
\section{Introduction}
People increasingly rely on a growing ecosystem of personal digital tools: health apps that log workouts and sleep, parking services that track time and payments, calendars, location traces, wearable sensors, and lightweight dashboards for everyday routines. Each tool is useful in isolation, yet everyday personal work rarely stays within a single application, and many data is interconnected. A simple situated question such as ``\textit{Is my heart rate too high right now?}'' may require combining recent workout activity, current motion, sleep quality, and even contextual signals such as whether the user is still walking back to an expiring parking spot.
% People increasingly rely on a growing ecosystem of personal digital tools: health apps that log workouts and sleep, parking services that track time and payments, calendars, wearable sensors, and lightweight dashboards for everyday routines. Yet everyday questions rarely stay inside one application. In the motivating episode in Figure~\ref{fig:teaser}, answering ``\textit{Is my heart rate too high right now?}'' requires combining recent workout activity, current motion, poor sleep, and whether the user is still walking back to an expiring parking spot. 
The difficulty is not that any single signal is unavailable; it is that these signals remain fragmented across apps, services, and interfaces.

This fragmentation is a long-standing problem in personal informatics~\cite{li2010stage, epstein2015lived, choe2014understanding}. Recent AI coding agents make it increasingly plausible for one person to generate lightweight, highly personalized tools from natural-language intent, but generation alone does not solve the integration problem. If each generated artifact becomes another siloed app, personal software scales in quantity rather than coherence~\cite{li2017sugilite, li2019pumice, ko2011state, nardi1993small}.

We present \textbf{PSI}, a shared-state architecture for coherent AI-generated instruments for AI-generated personal software. 
Borrowing the term from Beaudouin-Lafon's instrumental interaction model~\cite{beaudouinlafon2000instrumental} while extending it to AI-generated personal software, we define an \emph{instrument} as a generated artifact that is (a)~\emph{persistent}: it remains available without regeneration; (b)~\emph{connected}: it publishes state to a shared personal-context layer and may expose write-back affordances; and (c)~\emph{complementary to chat}: it supports glanceable monitoring while chat handles synthesis, ambiguity resolution, and stateful actions. A \emph{module} is the full software bundle behind an instrument, its GUI, provider, and optional services. Both persistent instruments and a generic chat agent (\textbf{Facai}) operate over the same shared-context bus. The result is not a new coding agent, but a minimal integration contract that lets independently generated modules become legible to one another and to multiple interfaces.

We study PSI through a three-week autobiographical deployment ~\cite{neustaedter2012autobiographical, desjardins2018living} in RyanHub, a self-developed personal AI environment, together with a broader artifact in which new modules can be generated and then automatically integrated into PSI through the same contract and registration path. The point is not the exact module count, but that independently authored modules can join the shared context after generation. Our aim is a versatile application rather than general utility: the deployment involves a single technically skilled user, and the evaluation is a bounded proof-of-concept rather than a claim about population-wide adoption.

This paper makes two contributions:
\begin{enumerate}
\item \textbf{A shared-state architecture for coherent AI-generated instruments} that turns independently generated modules into persistent, connected artifacts accessible through both GUI surfaces and chat.
    % \item \textbf{A provider contract and shared personal-context runtime} that turns independently generated modules into coherent instruments---persistent, connected, and accessible through both GUI surfaces and chat.
    \item  \textbf{Evidence that shared state improves both reasoning and action in personal AI}, showing stronger cross-module reasoning than search-only or single-module baselines while preserving reliable write-back across persistent instruments in an autobiographical deployment.
    \item \textbf{An open-sourced artifact} including the PSI, RyanHub IOS app, and representative modules, will be open-sourced upon acceptance.
    % \item \textbf{An instantiated artifact and proof-of-concept evaluation} showing that this contract supports cross-module reasoning and reliable write-back in an autobiographical deployment, while allowing later-generated modules to be integrated into the same architecture automatically.
\end{enumerate}

\section{Related Work}

\textbf{AI-generated software and end-user programming.} PSI builds on a long arc of end-user software creation, from task-specific end-user programming environments~\cite{nardi1993small} and end-user software engineering~\cite{ko2011state} to specification-by-demonstration systems such as SUGILITE and PUMICE~\cite{li2017sugilite, li2019pumice}. Commercial copilots and recent arguments for malleable software similarly shift attention toward software generation as an end-user-facing capability~\cite{copilot2021, litt2023malleable}. More recent research systems move closer to dynamic UI synthesis and agentic software production~\cite{vaithilingam2024dynavis, cao2025malleable, suh2024luminate, claude-code}. This literature shows that users can increasingly author or request new software artifacts; PSI focuses on a different question: what runtime contract makes many generated personal artifacts cohere after they have been created?

\textbf{Interactive AI interfaces beyond chat.} Direct-manipulation and post-WIMP traditions emphasize persistent, inspectable interaction objects rather than transient dialogue alone~\cite{shneiderman1983direct, norman2013design, beaudouinlafon2000instrumental}. Classic visions of ubiquitous and personally meaningful computing similarly foreground interfaces that remain embedded in everyday life rather than appearing only on demand~\cite{weiser1991computer}. Recent LLM interaction work shows the value of visual structure, diagrams, and hybrid interfaces~\cite{masson2024directgpt, xia2023graphologue, kim2021dataathand}. PSI adopts the intuition that chat should not be the only interface to personal AI, but pushes it toward an architectural claim: persistent GUI instruments and chat become complementary only when they operate over the same underlying state.

\textbf{Personal informatics, context, and proactive assistance.} Personal informatics research has long identified fragmentation and integration as core challenges~\cite{li2010stage, epstein2015lived, choe2014understanding}. Context-aware computing argues that richer state enables more appropriate assistance~\cite{dey2001understanding}, while proactive-assistant and mixed-initiative work highlights the long-standing appeal of systems that reduce information and coordination burden without removing the user from the loop~\cite{maes1994agents, horvitz1999mixed}. Recent AI systems such as OmniActions and GLOSS derive assistance from multimodal sensing and language models~\cite{grossman2024omniactions, choube2025gloss}. PSI differs in publishing person-scoped, module-produced state that persists across sessions and is shared by both conversational and graphical interfaces, rather than only supporting immediate prediction or one-shot interpretation.

\textbf{Method, architectural substrates, and agent memory.} Our evidence comes from an autobiographical deployment, a method for systems whose value depends on authentic everyday use~\cite{neustaedter2012autobiographical, desjardins2018living}. This also aligns with cultural-probe, technology-probe, and research-through-design traditions that use artifacts to surface design knowledge and system tensions~\cite{gaver1999design, hutchinson2003technology, zimmerman2007rtd, gaver2012rtd}. At the architectural level, PSI is closest to interaction substrates~\cite{mackay2025substrates}: it offers a reusable integration layer rather than a single application. At the agent level, prior work on generative agents, personal LLM agents, user modeling, and personal knowledge ecosystems similarly seeks person-relevant state~\cite{park2023generative, li2024personal, shaikh2025gum, zhao2025knoll}, but typically keeps that state inside an agent or knowledge model rather than exposing it as a shared runtime contract for multiple interfaces.

\section{PSI System Overview}

\begin{table}[t]
\caption{Capability comparison for personal AI systems. \checkmark~= full, $\sim$~= partial, --~= not supported. \emph{Conv.}: conversational AI; \emph{Gen.}: generates persistent GUIs; \emph{Ctx}: structured personal context; \emph{X-Mod}: cross-module synthesis; \emph{App}: app-level write-back.}
\label{tab:comparison}
\small
\begin{tabular}{lccccc}
\toprule
 & \rotatebox{60}{\small Conv.} & \rotatebox{60}{\small Gen.} & \rotatebox{60}{\small Ctx} & \rotatebox{60}{\small X-Mod} & \rotatebox{60}{\small App} \\
\midrule
ChatGPT / Claude & \checkmark & $\sim$ & $\sim$ & -- & -- \\
Siri / Google Asst. & \checkmark & -- & $\sim$ & -- & -- \\
Shortcuts / IFTTT & -- & -- & -- & $\sim$ & $\sim$ \\
Home Assistant & $\sim$ & -- & \checkmark & \checkmark & \checkmark \\
v0 / Lovable & \checkmark & \checkmark & -- & -- & $\sim$ \\
M365 Copilot & \checkmark & -- & $\sim$ & \checkmark & $\sim$ \\
Copilot / Cursor & \checkmark & -- & -- & -- & -- \\
DynaVis~\cite{vaithilingam2024dynavis} & -- & $\sim$ & -- & -- & -- \\
SUGILITE~\cite{li2017sugilite} & $\sim$ & -- & -- & $\sim$ & $\sim$ \\
OmniActions~\cite{grossman2024omniactions} & \checkmark & -- & \checkmark & -- & -- \\
OpenClaw~\cite{openclaw2025} & \checkmark & $\sim$ & $\sim$ & $\sim$ & $\sim$ \\
\midrule
\textbf{PSI} & \checkmark & \checkmark & \checkmark & \checkmark & \checkmark \\
\bottomrule
\end{tabular}
\end{table}

\begin{figure*}[t]
\centering
\includegraphics[width=\textwidth]{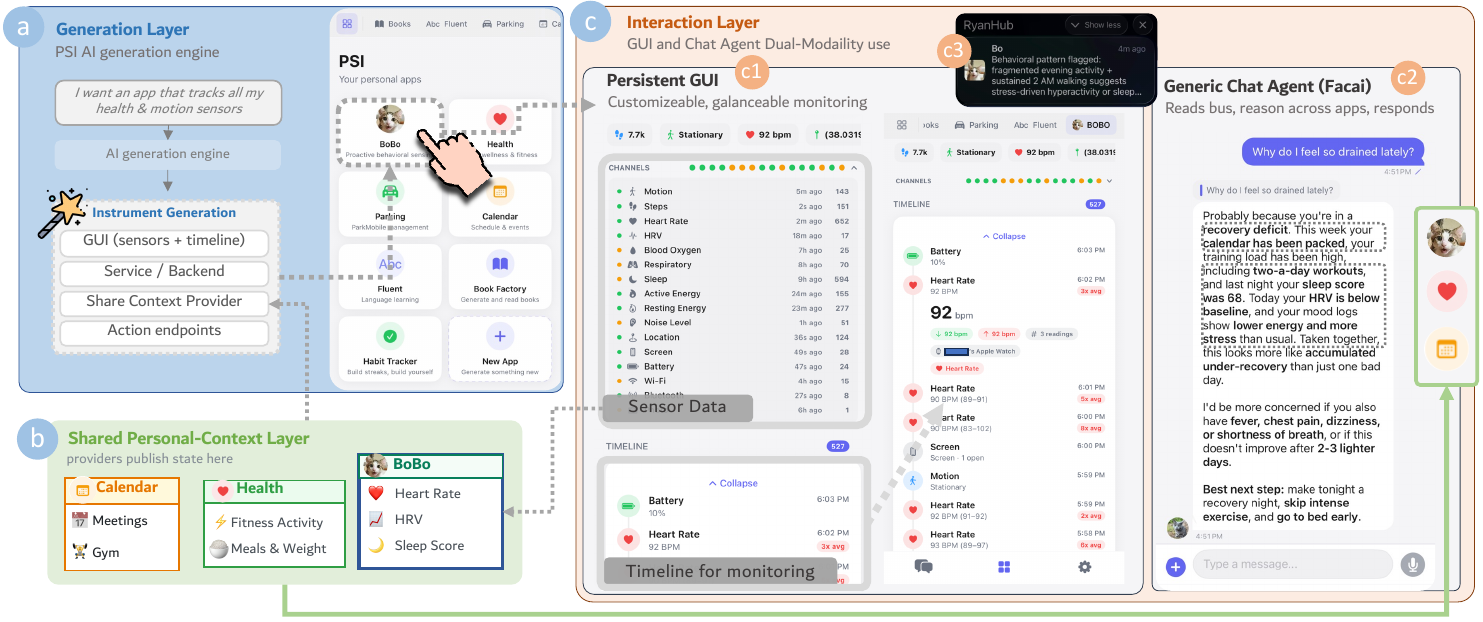}
\caption{PSI pipeline and interface walkthrough: PSI turns generated personal apps into persistent, connected, and chat-complementary instruments.
(a) Ryan uses PSI to generate BoBo, a personalized health instrument that connects passive sensor streams such as motion, steps, heart rate, and sleep.
(b) BoBo publishes its state to PSI’s shared personal-context bus, enabling interoperability with other instruments and apps (e.g., calendar and health logs).
(c1) PSI provides a persistent, customizable GUI with a glanceable dashboard and interactive timeline for longitudinal monitoring.
(c2) When Ryan asks Facai, ``Why do I feel so drained lately?'', the agent retrieves relevant state across connected instruments and jointly reasons over sleep, activity, and calendar load to provide a grounded explanation without requiring Ryan to manually inspect fragmented apps. (c3) BoBo will actively track the user status and nudge the user proactively. }
\label{fig:pipeline}
\end{figure*}

% No existing system combines all five capabilities. PSI's contribution is the architectural contract, the provider protocol, that achieves their conjunction through a shared personal-context layer.
Prior systems support pieces of the personal AI workflow---generation, sensing, automation, or conversational access---but the experience breaks down after creation because these capabilities remain isolated. PSI introduces \emph{instruments}: persistent, connected, chat- complementary artifacts that address this missing layer through a shared personal-context substrate and provider contract, letting independently created modules interoperate and remain accessible through both GUI surfaces and chat.

\subsection{Motivating Scenarios}

Ryan has recently been having trouble making sense of his passive health data. Signals such as motion, step count, heart rate, sleep, and other sensor streams are continuously collected, but the information remains fragmented across separate apps and logs.

With PSI, he quickly generates a personalized modules (Figure~\ref{fig:pipeline}a) called \textbf{BoBo} (Behavioral Observer Bot), which connects to all of his health- and motion-related sensors.

% BoBo provides two key benefits. 
First, Ryan wanted a customized timeline to track these sensor data. Hence, PSI helps Ryan to create a \textit{persistent, customizable GUI} (Figure~\ref{fig:pipeline}c1) specific for BoBo that visualizes these signals along an interactive timeline, allowing Ryan to monitor trends over time at a glance.
Second, it publishes its state to PSI's \textit{shared personal-context bus} (Figure~\ref{fig:pipeline}b), allowing the app to interoperate with other existing instruments, such as health logging and parking history.                 

At the center of this ecosystem is \textbf{Facai}, 
a \textit{generic chat agent} (\autoref{fig:pipeline}c2) that can interpret user questions, coordinate across modules, and reason over the shared personal-context bus, and nudge user proactively (\autoref{fig:pipeline}c3).

One day, while walking between meetings, Ryan suddenly notices that his heart is racing, and he felt so tired lately. 
Unsure whether this is normal or a sign of something concerning, he asks Facai: ``Why do I feel so drained lately?''
(Figure~\ref{fig:pipeline}c2).

Instead of forcing Ryan to manually open and compare multiple disconnected apps, Facai retrieves relevant state from BoBo, his other modules, such as Health, and even Calendar history through the shared context bus. By jointly reasoning over these connected signals, Facai helps Ryan discover the likely explanation: he had been rushing from meetings and interviews, exercising too much,  with poor sleep the night before. 

Through PSI, Ryan no longer needs to inspect fragmented apps one by one. Instead, PSI turns generated apps into \textit{instruments}: persistent, connected, and chat-complementary artifacts that support both glanceable monitoring through GUIs and cross-context reasoning through conversation.

% Ryan has some problem with analyzing his passive health data lately, where proactive passive data such as motion, steps, heart rate and sleep, and ...
% Using PSI, he quickly creates a module, named BoBo (Behavioral Observer Bot) that help access to all the health and motion related sensor. 

% This module has two benefits,  first it creates and customizes a persistent GUI app for tracking these data on a timeline. Second, this module can also connects to all the other existing related modules, such as health and parking with a shared personal context bus. A \textit{generic chat agent} named \textbf{Facai}, can handle the user's question and manages the shared personal context bus and achieving any relevant data from the shared context bus. 

% % - health related sensor...
% % - health behavior: 

% % Health: weight, food, activity: user actively logging data -----

% One day, when Ryan is walking on the go, where he felt his heart is beeping. He doesn't know why this happened, and how this can be a problem. 
% hence， he asked \textbf{Facai}, the generic chat agent that handles every module he has authored， ``Isn't my current
% heart rate too high?''. Facai then looked into the data, and find the data from BoBo, Health, and Parking. 

% This way, Ryan were able to find the context and data rather than check every fragmented app he has. 

\subsection{PSI Architecture}
PSI consists of three layers: a generation layer for authoring modules, a runtime layer built on the shared personal-context bus, and an interaction layer that exposes module state through both persistent instruments and a chat agent.
% PSI consists of three layers.
% % (Figure~\ref{fig:architecture})
% The generation layer uses an AI coding agent to produce and revise modules from natural-language requests. The runtime layer is a shared personal-context bus. The interaction layer exposes the same underlying module state through both a chat agent (Facai) and persistent GUIs (e.g., timeline, calendar).
\subsubsection{Generation Layer}

PSI enables an agentic coding workflow rather than a fixed application catalog. Modules are generated through a multi-phase pipeline, consisting of specification, code generation, auto-fix, and compile verification. 
The architectural contribution is not the pipeline itself but the provider contract each generated module must satisfy. The formal contract is a single Swift protocol (\texttt{ToolkitDataProvider}) requiring a toolkit identifier, relevance keywords, and one method---\texttt{buildContextSummary() -> String?}---that returns a tagged, human-readable snapshot of the module's current state (e.g., today's sensed events, recent meals, upcoming calendar entries) together with any write-back endpoints it exposes, so the same method serves both read context and action discovery for the chat agent. A co-evolving memory file captures informal conventions (naming, data formats) discovered during development but not enforced at compile time; this is how modules generated on different days fit one architecture.

\subsubsection{Shared Personal Context Layer}

The shared personal-context layer is a central registry that collects current-state snapshots from all registered modules---both built-in and dynamically generated---and prepends them as a single tagged block before every chat message. Each snapshot captures today's data rather than cumulative history; if a module has nothing to report it is silently omitted, so the system degrades gracefully. This turns integration into a local obligation: a new module needs only to implement the provider interface and register, rather than wire into every existing module. Unlike agent orchestration frameworks where state is scoped to a single task, PSI's shared context is person-scoped---it persists on-device across tasks and sessions, requires no coordinating task graph, and is consumed by all interfaces. Modules never read each other's state directly; all cross-module communication is mediated by the LLM through the assembled context.

\subsubsection{Interaction Layer: Dual-Modality Use}
The interaction layer provides two coordinated interfaces: a persistent GUI (\autoref{fig:pipeline}c1) and a generic chat agent (\autoref{fig:pipeline}c2). PSI’s contribution is not simply the coexistence of chat and GUI, but their role as synchronized entry points to the same person-scoped mutable state. Users can inspect state in a persistent instrument, revise it through chat, and immediately verify the effect in the GUI without duplicated state paths or re-specifying intent. For example, in BoBo, the behavioral timeline remains persistently glanceable in the GUI while Facai reasons over the same visible state in follow-up queries. 
\section{Evidence}
\label{sec:evidence}

We evaluate PSI through two complementary evidence lenses~\cite{olsen2007evaluating,greenberg2008usability,ledo2018evaluation}: 
Applications and Proof-of-concept.
% \textbf{architectural reach through a minimal integration surface} and \textbf{scalability and performance} (tractability at current scale with a proof-of-concept evaluation).

For evaluation, the PSI system is instantiated in RyanHub, a self-developed personal AI environment comprising four cooperating services: (1)~a SwiftUI iOS app (94 Swift files, 36{,}517 LOC) hosting persistent GUI instruments and the chat interface;
% ; six built-in instruments are used in the autobiographical deployment: BoBo, Health, Calendar, Parking, BookFactory, and Fluent. The same architecture now also hosts eight dynamically generated modules (Dashboard, Habit Tracker, Hydration Tracker, Mood Journal, Spending Tracker, etc.), showing that new modules can be generated and then automatically integrated into PSI through the same provider contract and registration path
(2)~a Python bridge server providing a unified REST gateway for on-device data (behavioral timeline, health entries, parking state) persisted as local JSON files; (3)~a Python dispatcher maintaining WebSocket sessions with the iOS client, injecting shared context server-side, and routing tool calls through the LLM;
% and (4)~a \textbf{Next.js service} for BookFactory module (generative audiobook management with SQLite storage). 
All services run on localhost; personal data stays on-device by default.

% For evaluation, PSI is instantiated in \textbf{RyanHub}, a SwiftUI iOS app backed by a Python bridge server, a Python dispatcher, and a Next.js book service. The current artifact contains 94 Swift files (36{,}517 LOC) and six built-in instruments used in the autobiographical deployment: BoBo, Health, Calendar, Parking, BookFactory, and Fluent. The same architecture now also hosts eight dynamically generated modules (Dashboard, Habit Tracker, Hydration Tracker, Mood Journal, Spending Tracker, etc.), showing that new modules can be generated and then automatically integrated into PSI through the same provider contract and registration path. Persistent GUI instruments are organized through a server-backed personal toolkit shell, while the generic chat agent Facai persists across tab switches and reads the same person-scoped state.

\subsection{Versatile Applications}
% ity Through a Minimal Integration Surface}
\label{sec:versatility}

% PSI's 
% strongest evidence is architectural: 

PSI supports one shared personal-context contract supports a heterogeneous tool ecosystem without pairwise integration. We created 
% The current artifact comprises 
14 modules across behavioral sensing, health, scheduling, parking, reading, vocabulary learning, and several post-pilot self-tracking domains. 
These include six core modules used during the three-week autobiographical deployment, along with eight additional modules newly generated to validate the extensibility of the generation layer (see the full listing in Appendix~\ref{app:modules}). 
% Despite being created after the pilot, these later modules were 
% integrated automatically through the same contract and registration path. All publish shared personal- context and expose persistent GUIs, and also expose chat-invocable write-back or trigger pathways. 

% \subsubsection{Artifact Implementation}

% A full listing appears in Appendix~\ref{app:modules}.

% This breadth is enabled by a deliberately small integration layer spanning 173 lines across four files: the provider protocol, context assembly runtime, dynamic registry, and startup bootstrap. Built-in modules join by implementing \texttt{ToolkitDataProvider}; dynamically generated modules additionally register one descriptor. Chat consumes shared state through a single call to \texttt{PersonalContext. buildContext(...)} before message dispatch, and the toolkit home screen renders dynamic modules from the same registry. The key shift is not zero-cost integration but localized integration: new modules no longer require direct chat wiring or pairwise adapters into existing ones. 

Here, we present two deployment cases illustrate the payoff of that contract in everyday use, Bobo and Automated Parking. 

% \begin{figure}[t]
% \centering
% \includegraphics[width=\columnwidth]{figures/figure2-glance-ask-act.pdf}
% \caption{The \emph{glance--ask--act} pattern. \textbf{Left:} the user glances at BoBo's behavioral timeline. \textbf{Middle:} a follow-up chat query. \textbf{Right:} a grounded response synthesizing health, behavioral, and scheduling state from the same shared context.}
% \Description{Three-panel figure showing the glance-ask-act interaction: BoBo timeline, chat query, and cross-module response.}
% \label{fig:glance-ask-act}
% \end{figure}

\begin{figure}[t]
\centering
\includegraphics[width=0.7\columnwidth]{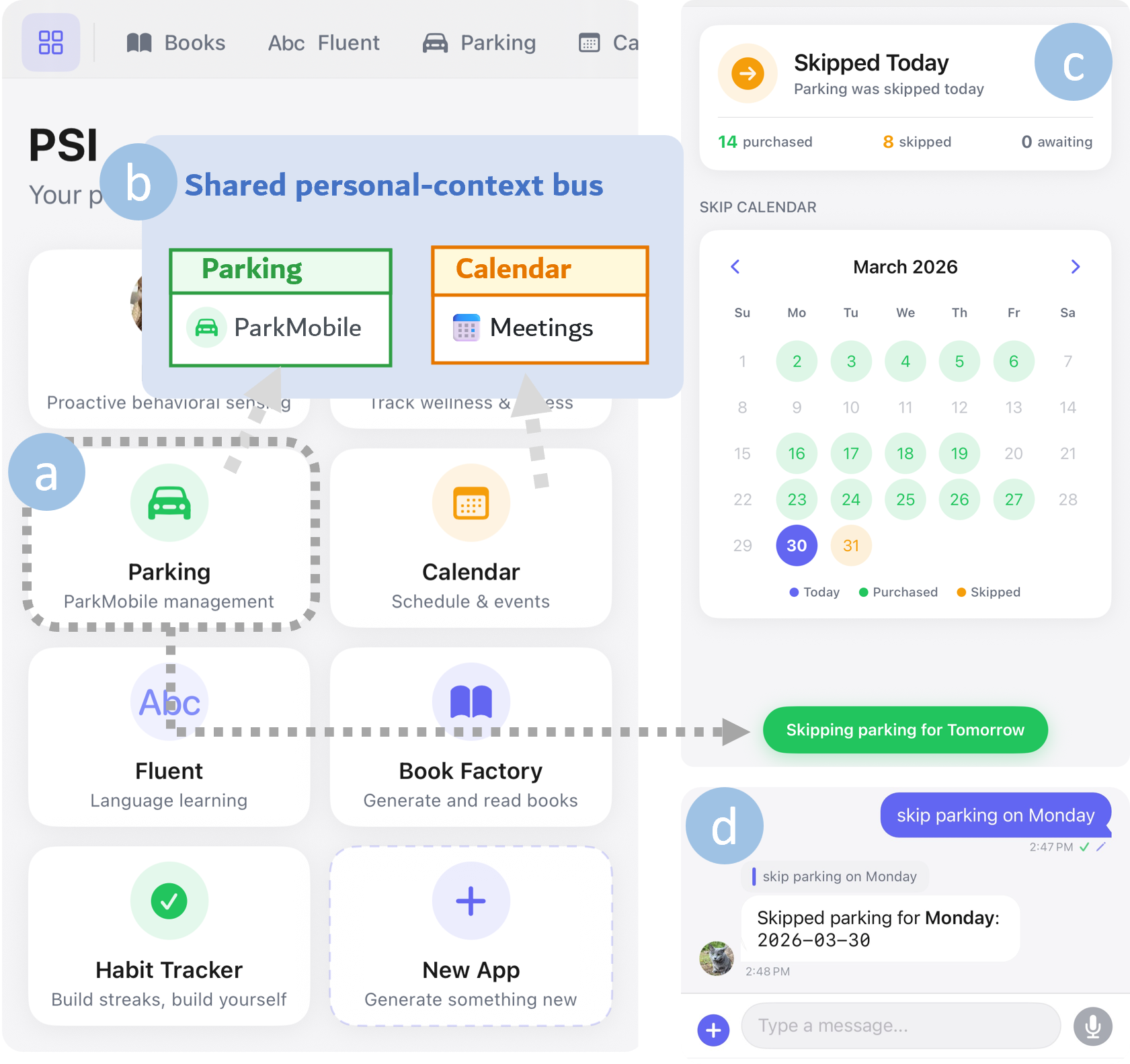}
\caption{Automated Parking Example}
\label{fig:parking}
\end{figure}

% \paragraph{BoBo: a Proactive Behavioral Sensing Module}
% \paragraph{\red{From Glance to Action (Figure~\ref{fig:glance-ask-act}).}}
\paragraph{BoBo: A Generalizable Behavioral Sensing Instrument.}
Beyond the motivating walkthrough, BoBo demonstrates how PSI supports a broader class of persistent behavioral sensing instruments. Rather than serving a single question-answer interaction, BoBo maintains a continuously updated behavioral state that can be accessed through both glanceable GUIs and conversational reasoning. This enables diverse query patterns, including cross-signal synthesis (e.g., relating heart rate spikes to location, activity, and calendar load), temporal grounding (e.g., comparing sleep or recovery trends over multiple days), and action-oriented follow-ups (e.g., recommending rest, suppressing evening workouts, or adjusting the next day’s schedule). Because these interactions operate over the same shared personal-context substrate, users can fluidly move between passive monitoring, situated questioning, and proactive intervention without manually reconstructing state across fragmented apps.
% As shown in Figure~\ref{fig:pipeline}, BoBo illustrates how a persistent instrument and chat can share one person-scoped substrate. The user first glances at the timeline directly in the GUI, then asks a follow-up question in chat. Because Facai reads the same shared context, it can ground its response in the already-visible behavioral state rather than reconstructing the day through ad hoc search. Furthemore, Facai can proactively intervent and proactively nudge the user...
% The value is not the answer alone, but that the instrument makes state persistently visible while the chat agent remains synchronized with it.

% \paragraph{Market-of-One Parking.}
\paragraph{Automated Parking}
Ryan faces a recurring parking challenge: if he does not reserve a spot before 7 a.m., the lot is typically fully booked. To avoid waking up early for this routine, he created a module that automatically books parking on his behalf.
The Parking module demonstrates the same PSI pattern in a hyper-personal, \emph{market-of-one} workflow. Tailored to a single user’s weekday parking routine, it supports configurable zones, vehicles, and schedules. Through PSI, the generic chat agent Facai can trigger ParkMobile purchases via web automation, while the user may also interact through a persistent GUI. For example, the user can issue a command such as \textit{“No parking this Thursday”} (Figure~\ref{fig:parking}d) or toggle the same skip state directly in the GUI (Figure~\ref{fig:parking}c). The module can also integrate state from other instruments via the shared personal-context bus (Figure~\ref{fig:parking}b), such as using the calendar’s end-of-day event to infer parking duration.
% The Parking module shows the same pattern in a hyper-personal workflow.
% Rather than serving as a generic trip planner, it functions as a \emph{market-of-one} automation tailored to a single user’s weekday parking routine, with configurable parking zones, vehicles, and schedules. With PSI, a generic chat agent, \textbf{Facai}, can trigger ParkMobile purchases through web automation, while the user may also interact through a persistent GUI. For example, the user can issue a chat command such as \textit{“No parking this Thursday”} (Figure~\ref{fig:parking}d) or directly toggle the same skip state in the GUI (Figure~\ref{fig:parking}c). 
% Furthermore, the parking automation can read from other modules through the shared personal-context bus (Figure~\ref{fig:parking}b), e.g., using the end of the user’s workday from the calendar module to determine the appropriate parking duration.
In both modes, chat and GUI operate over the same persistent parking state, including schedule, purchase history, and active sessions, as a single source of truth.
More broadly, this pattern generalizes to other recurring \emph{market-of-one} routines, such as gym bookings, commute ticketing, medication reminders, and home-device schedules.

\subsection{Proof-of-concept Evaluation}
\label{sec:scalability}

\begin{figure}[t]
\centering
\begin{minipage}[t]{0.28\columnwidth}
    \centering
    \includegraphics[width=\linewidth]{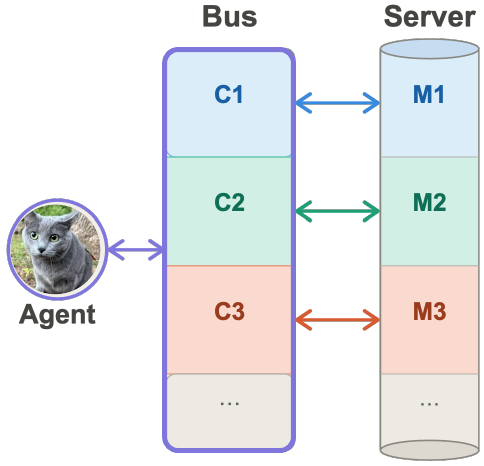}\\
    \small (a) Context Bus  
    \label{fig:cb}
\end{minipage}\hfill
\begin{minipage}[t]{0.28\columnwidth}
    \centering
    \includegraphics[width=\linewidth]{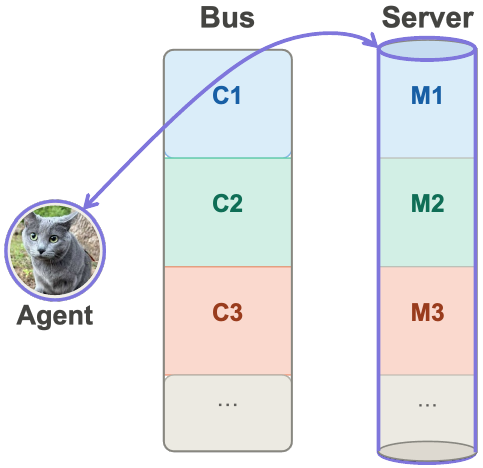}\\
    \small (b) Search-Only
    \label{fig:so}
\end{minipage}\hfill
\begin{minipage}[t]{0.28\columnwidth}
    \centering
    \includegraphics[width=\linewidth]{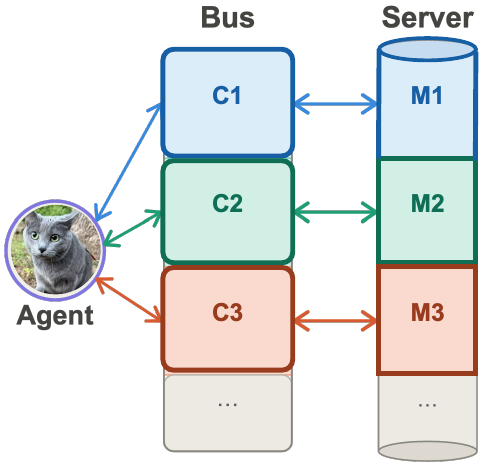}\\
    \small (c) Single-Module
    \label{fig:sm}
\end{minipage}
\caption{Benchmark illustrations. C = Context; M = Module.}
\label{fig:bench-panels}
\end{figure}

% \paragraph{Prompt-size growth.}
% Shared context grows linearly with module count. Across six deployed modules, provider summary averaged approximately ${\sim}$200 tokens per module, producing 1{,}223 tokens in total.  At 50 modules this would reach ${\sim}$10{,}000 tokens, within current context windows but approaching the point where attention degradation may reduce reasoning quality. The expected scaling bottleneck is LLM attention over accumulated context, not assembly cost.
% We treat this evaluation as suggestive, not definitive. 
% \paragraph{Proof-of-concept evaluation.}
To understand the benefit and performance of generic chat agent powered by shared-personal context bus across modules, 
we evaluated three conditions: (1) Shared Personal-Context; (2) Search-Only; (3) Single-Module (\autoref{fig:bench-panels}a-c). 
In \emph{Search-Only}, the agent received no preassembled personal snapshot and had to recover relevant state opportunistically from user database from in the file system during the turn (comparable to how OpenClaw works). In \emph{Single-Module}, we ran the same task once per candidate module, and report the best-rated one-module variant for each condition.

\paragraph{Data Collection Tasks}
We evaluated the generic chat agent, powered by the shared personal-context bus, using a frozen three-week dataset collected in a single-user, self-authored deployment setting. The dataset included both proactively logged personal data (e.g., food intake and diary entries) and passively sensed data streams from a smartwatch, such as heart rate, location, and ambient noise levels.
We then generated a set of synthetic user queries (N=50) at different time points to mimic real-world use, where the system has access to only the data available at that moment. 
 
The 50 evaluation queries were organized into three \textbf{reasoning} task categories commonly studied in prior work: \emph{cross-module synthesis}~\cite{li2010stage,epstein2015lived,dey2001understanding,kim2021dataathand,choube2025gloss} (e.g., ``Given my today's done, what is the best next step for tonight?''), \emph{temporal grounding and control}~\cite{li2010stage,epstein2015lived,choe2014understanding} (e.g., ``How has my heart rate changed over the last week?''), and \emph{chain of actions}~\cite{li2017sugilite,li2019pumice,grossman2024omniactions} (e.g., ``Check my calorie intake, then activity, then net balance.''). 
The distribution of simulated queries across these categories was derived from the questions the single user asked during the three-week deployment.
Aside from reasoning tasks, we tested 20 \textbf{write-back} action tasks across five domains (parking, food, activity, diary, and dynamically generated modules), to assess whether state changes initiated from chat were correctly reflected in the corresponding GUIs, as validated by sandbox state inspection.
% The full task set and per-task results are available as supplementary material.

\paragraph{Metrics and Results}
We evaluated the resulting responses using \textit{fulfillment}, \textit{task success}, and \textit{latency}.
\emph{Fulfillment} measures the fraction of gold-specification criteria satisfied by a response (continuous, 0--1), operationalized as the proportion of relevant modules correctly identified among all ground-truth modules. \emph{Task success} is a stricter binary metric that requires all relevant modules to be selected (0 or 1).
Because these tasks require integrating longitudinal evidence across many modules and time windows, holistic human rating would itself require reconstructing fragmented personal traces, a burden that personal informatics research has long identified as difficult in practice~\cite{li2010stage,epstein2015lived,choe2014understanding}. We therefore use an independent language-model judge (Claude Opus 4.6) as a pragmatic proxy rather than a substitute for human evaluation.
% In all the three conditions, we use the same underlying LLM (GPT-5.4 via Codex); 
% % \red{the only independent variable is the information architecture (shared context injection vs.\ single-module context vs.\ ad hoc file search).} 
% Reasoning criteria were evaluated by an independent language model judge (Claude Haiku), \red{which has been shown to correlate well with human ratings on factual grounding tasks. }
% \red{Three conditions were compared: \emph{Shared Context} ($N$=50, full six-module injection), \emph{Search-Only} ($N$=50, no injection; agent may recover state ad hoc from local files), and \emph{Single-Module} ($N$=49, best-scoring single-module variant per task).} 
% Table~\ref{tab:benchmark-results} reports the reasoning results. 
% \emph{Fulfillment} measures the fraction of gold-specification criteria a response satisfies (0--1 continuous) based on how many modules are correctly identified across all the true modules; \emph{task success} requires all criteria met (binary: 0 or 1). 

Shared personal-context achieved a mean \textit{fulfillment} score of 0.88, substantially outperforming Search-Only (0.63) and Single-Module (0.27). \textit{Task success} followed the same pattern at 0.68, 0.32, and 0.08, respectively. We also measured latency. End-to-end latency is not monotonic with context size. On reasoning tasks, the mean successful latency was 25\,s for Shared Context, 29\,s for Search-Only, and 23\,s for Single-Module. 

On write-back actions, we measure the \textit{task success} on whether the GUI task were precisely completed by the chat agent or not. 
The pattern reverses: Shared and Single-Module both achieved 19 of 20 validated state changes (95\%), while Search-Only achieved 8 of 20 (40\%). 

% Shared Context achieved 0.88 mean fulfillment versus 0.63 for Search-Only and 0.27 for Single-Module; task success is 0.68 versus 0.32 versus 0.08. By task family: \emph{cross-module synthesis} (34~tasks) showed 0.91 vs.\ 0.66---the bus enables simultaneous multi-instrument reads that search cannot guarantee; \emph{temporal grounding} (10~tasks) showed 0.70 vs.\ 0.45---the injected snapshot pins the model to a specific time window that ad hoc search cannot reconstruct.

% This is expected---each action targets a single module, so the matching module's context suffices for routing. 

In summary, the gap between conditions reveals where shared context's value lies: for \emph{reasoning}, it enables cross-module synthesis that no single module can provide; for \emph{actions}, it provides write-path discovery that ad hoc search cannot reliably achieve. Together, the reasoning and action results provide evidence for \emph{bidirectional information access}: shared context lets chat both read personal state (grounded reasoning) and write it back (stateful actions).

% \begin{table}[t]
% \caption{Reasoning evaluation ($N$=50 per condition). Scored by independent LLM judge. Single-Module reports the best-scoring variant per task.}
% \label{tab:benchmark-results}
% \centering
% \small
% \begin{tabular}{lccc}
% \toprule
%  & \textbf{Shared} & \textbf{Search} & \textbf{Single} \\
% \midrule
% Fulfillment & 0.88 & 0.63 & 0.27 \\
% Task success & 0.68 & 0.32 & 0.08 \\
% Avg.\ latency & 25\,s & 29\,s & 23\,s \\
% \bottomrule
% \end{tabular}
% \end{table}

% \paragraph{Latency.}

% ============================================================
\section{Discussion and Conclusion}

\textbf{From generated apps to coherent instruments.} The deployment suggests that the value of PSI lies not in any single module but in the division of labor that instruments enable. Instruments supported glanceable monitoring and routine control---the BoBo timeline and parking controls were often useful without opening chat at all---while chat handled cross-module synthesis and stateful actions over the same shared state. Without the shared-context layer, each generated module would remain a local success but a system-level dead end; the provider contract is what turns isolated apps into instruments by making integration a local obligation rather than a pairwise problem.

% \textbf{Why the runtime contract matters.} PSI's main contribution is a small integration contract, not a comprehensive personal AI platform. Without a shared-state layer, each generated personal module remains a local success but a system-level dead end. With the intrument, adding a new module becomes a local obligation---implement a summary and register it---instead of a pairwise integration problem. This locality is what lets RyanHub combine built-in modules and generated modules in one coherent environment.

\textbf{Current design bets and limitations.} Unconditional injection trades prompt length for recall. At current scale the injected context is still manageable, but routing and selection will matter more as module counts grow. The deployment also surfaces a system-specific risk: \emph{context pollution} \cite{liu2024lost}. Because all interfaces trust provider summaries, stale or misleading summaries can degrade responses system-wide. As the number of modules grows, overlapping or redundant entries logged across modules can introduce ambiguity into the assembled context, causing the chat agent to misattribute, double-count, or contradict itself, a class of failure inherent to shared-state architectures that would not arise in siloed apps. 

% PSI currently offers no automatic reconciliation, provenance checking, or conflict resolution across modules.

\textbf{Future work.} Our evidence comes from a single technically skilled user over three weeks, so the paper should be read as a proof-of-concept for versatile applications. The reusable instrument that includes persistent GUI and generic chat agent are intended to be user-agnostic, but the current summaries, modules, and deployment practices are specific to one personal ecology. Important open questions include routing at larger module counts, privacy and authorization for person-scoped actions, and how to make compliant module generation accessible to non-programmers. We also plan to release the reusable artifact components after publication, which should make these follow-on questions easier to study.

% ============================================================
% \section{Conclusion}

% PSI argues that the core systems problem in AI-agent-generated personal software is not generation alone, but integration after generation. By giving generated modules a shared personal-context runtime, PSI lets a generic chat agent and persistent GUIs operate over the same person-scoped state. In our proof of concept, this contract is lightweight enough to let newly generated modules be integrated automatically alongside built-in ones, yet consequential enough to improve cross-module reasoning and reliable write-back relative to weaker baselines. The architecture suggests a lightweight shared-state contract can make AI-generated personal tools more coherent, and that coherence is worth studying as a systems and interaction-design problem in its own right.

\bibliographystyle{ACM-Reference-Format}
\bibliography{references-short}

% ============================================================
% APPENDIX (does not count toward 5-page limit)
% ============================================================
\appendix

\section{Evaluation Task Set}
\label{app:tasks}

The evaluation comprises 50 reasoning tasks across three families (cross-module synthesis, temporal grounding, and multi-step chains) and 20 write-back action tasks across five domains (parking, food, activity, diary, and dynamically generated modules). The full task set with queries, gold-specification criteria, and per-task scores is available as supplementary material. Table~\ref{tab:action-tasks} lists representative action tasks.

\begin{table}[H]
\caption{Action tasks ($N$=20, representative subset). Validated by sandbox state inspection. Shared = 19/20, Single = 19/20, Search = 8/20.}
\label{tab:action-tasks}
\centering
\small
\begin{tabular}{p{0.40\columnwidth}p{0.22\columnwidth}cc}
\toprule
\textbf{Query} & \textbf{Domain} & \textbf{Sh.} & \textbf{Se.} \\
\midrule
Skip parking for tomorrow. & Parking & \checkmark & -- \\
Actually restore parking for tomorrow. & Parking & \checkmark & \checkmark \\
Skip parking for all of next week. & Parking & \checkmark & \checkmark \\
Log lunch: egg and chicken curry with rice. & Health (food) & \checkmark & -- \\
Log a 30 minute run, 300 cal. & Health (activity) & \checkmark & -- \\
Log my weight: 87.5 kg. & Health (weight) & \checkmark & -- \\
Add a diary entry: great workout today. & Diary & \checkmark & \checkmark \\
Log 8 glasses of water today. & Dynamic module & \checkmark & \checkmark \\
I slept 7.5 hours, quality good. & Dynamic module & \checkmark & \checkmark \\
Track vitamin D this morning. & Dynamic module & \checkmark & \checkmark \\
\bottomrule
\end{tabular}
\end{table}

\section{Per-Task Results}
\label{app:per-task}

The full evaluation set with per-task scores across all conditions is available as supplementary material. Table~\ref{tab:example-tasks} shows representative examples from each task family.

\begin{table}[H]
\caption{Example tasks from each family with Shared Context fulfillment scores.}
\label{tab:example-tasks}
\centering
\small
\begin{tabular}{p{0.50\columnwidth}p{0.12\columnwidth}p{0.12\columnwidth}}
\toprule
\textbf{Query} & \textbf{Family} & \textbf{Ful.} \\
\midrule
Given my day so far, what is the best next step? & Synth. & 1.00 \\
My day was noisy and I hit the gym---evening plan? & Synth. & 0.67 \\
What do I have tomorrow afternoon? & Temp. & 1.00 \\
Will parking auto-purchase next week? & Temp. & 0.50 \\
Check calorie intake, then activity, then net balance. & Chain & 1.00 \\
Look for stress signs, then food, then rest vs.\ exercise. & Chain & 0.75 \\
\bottomrule
\end{tabular}
\end{table}

\section{Module Coverage}
\label{app:modules}

Table~\ref{tab:module-coverage} lists all modules deployed during and after the pilot period, with their integration surface.

\begin{table}[H]
\caption{Module coverage. \emph{Pilot} = generated during the three-week pilot; \emph{Post} = generated afterward using the same pipeline. Ctx = publishes shared context; GUI = persistent GUI; Write = chat-invocable write-back.}
\label{tab:module-coverage}
\centering
\small
\begin{tabular}{llccc}
\toprule
\textbf{Module} & \textbf{Period} & \textbf{Ctx} & \textbf{GUI} & \textbf{Write} \\
\midrule
BoBo (behavioral sensing) & Pilot &\checkmark & \checkmark & \checkmark \\
Health (food/activity) & Pilot &\checkmark & \checkmark & \checkmark \\
Calendar (scheduling) & Pilot &\checkmark & \checkmark & \checkmark \\
Parking (automation) & Pilot &\checkmark & \checkmark & \checkmark \\
BookFactory (reading) & Pilot &\checkmark & \checkmark & \checkmark \\
Fluent (vocabulary) & Pilot &\checkmark & \checkmark & \checkmark \\
\midrule
Sleep Tracker & Post & \checkmark & \checkmark & \checkmark \\
Medication Tracker & Post & \checkmark & \checkmark & \checkmark \\
Spending Tracker & Post & \checkmark & \checkmark & \checkmark \\
Mood Journal & Post & \checkmark & \checkmark & \checkmark \\
Hydration Tracker & Post & \checkmark & \checkmark & \checkmark \\
Habit Tracker & Post & \checkmark & \checkmark & \checkmark \\
Reading Tracker & Post & \checkmark & \checkmark & \checkmark \\
Dashboard & Post & \checkmark & \checkmark & \checkmark \\
\bottomrule
\end{tabular}
\end{table}

\section{Integration Surface}
\label{app:integration}

The core shared-context mechanism spans 173 lines across four files:

\begin{itemize}
    \item \textbf{Provider protocol} (38 lines): defines \texttt{ToolkitDataProvider} with \texttt{buildContextSummary() -> String?} and relevance keywords.
    \item \textbf{Context assembly} (60 lines): iterates over all registered providers, concatenates summaries inside \texttt{[Personal Context]} delimiters, and prepends to chat messages.
    \item \textbf{Dynamic registry} (59 lines): maintains an in-memory dictionary of \texttt{DynamicModuleDescriptor} entries, each holding a view builder and provider type.
    \item \textbf{Bootstrap} (16 lines): calls each module's registration function at app startup.
\end{itemize}

A provider summary follows a simple tagged format:
\begin{verbatim}
[Health Data]
Today: 1030 cal, 62g protein
Gym: 12 min, 65 cal burned
[End Health Data]
\end{verbatim}

\end{document}